\documentclass[12pt,preprint]{aastex}

\newcommand{\kms}{\>{\rm km}\,{\rm s}^{-1}}
\newcommand{\kmskpc}{\>{\rm km}\,{\rm s}^{-1}\,{\rm kpc}^{-1}}

\newcommand\degrees{^\circ}
\newcommand{\len}{a_{\rm B}}
\newcommand{\rbp}{R_{\rm B/P}}

\def\etal{{et al.}}

\slugcomment{Draft version on \today} 
\shorttitle{Confirmation of a Kinematic Diagnostic for Face-On B/P Bulges}
\shortauthors{M\'endez-Abreu \etal} 

\begin{document}   

\title{Confirmation of a Kinematic Diagnostic for Face-On Box/Peanut-Shaped Bulges\altaffilmark{1}}
\author{J. M\'endez-Abreu \altaffilmark{2,3,4}, E. M. Corsini\altaffilmark{3},
Victor P. Debattista\altaffilmark{5,6}, S. De Rijcke\altaffilmark{7}, 
J. A. L. Aguerri\altaffilmark{8}, \& A. Pizzella\altaffilmark{3}}

\altaffiltext{1}{Based on observations collected at the European  
Southern Observatory, Chile (ESO N$^o$ 76.B-0324(A)).}
\altaffiltext{2}{INAF-Osservatorio Astronomico di Padova, 
  vicolo dell'Osservatorio 5, I-35122 Padova, Italy; 
  jairo.mendez@oapd.inaf.it} 
\altaffiltext{3}{Dipartimento Astronomia, Universit\`a di Padova, 
  vicolo dell'Osservatorio 3, I-35122 Padova, Italy; 
  enricomaria.corsini@unipd.it;alessandro.pizzella@unipd.it}
\altaffiltext{4}{Universidad de La Laguna, Av. Astrof\'{\i}sico 
  Francisco S\'anchez s/n, E-38206 La Laguna, Spain}
\altaffiltext{5}{Centre for Astrophysics, University of Central Lancashire, 
  Preston, PR1 2HE, UK; 
  vpdebattista@uclan.ac.uk}
\altaffiltext{6}{RCUK Fellow} 
\altaffiltext{7}{Sterrenkundig Observatorium, Universiteit Gent, 
  Krijgslaan 281, S9, B-9000 Gent, Belgium; 
  sven.derijcke@ugent.be}
\altaffiltext{8}{Instituto de Astrof\'\i sica de Canarias, Calle Via L\'actea s/n,
  E-38200 La Laguna, Spain; 
  jalfonso@iac.es}


%
%

\begin{abstract} 

We present the results of high resolution absorption-line spectroscopy
of 3 face-on  galaxies, NGC~98, NGC~600, and NGC~1703  with the aim of
searching for box/peanut  (B/P)-shaped bulges. These observations test
and  confirm,  for  the   first,  time  the  prediction  that  face-on
B/P-shaped bulges can be recognized by two minima in the profile along
the bar's major  axis of the fourth Gauss-Hermite  moment $h_4$ of the
line-of-sight velocity distribution (LOSVD). In NGC~98, a clear double
minimum in $h_4$ is present along the major axis of the bar and before
the end  of the bar, as  predicted. In contrast, in  NGC~600, which is
also a barred galaxy but lacks a substantial bulge, we do not find any
significant kinematic  signature for a B/P-shaped  bulge. In NGC~1703,
which is  an unbarred control  galaxy, we found  no evidence of  a B/P
bulge.
We also show directly that the LOSVD is broader at the location of the
$h_4$ minimum in NGC~98 than elsewhere. This more direct method avoids
possible  artifacts   associated  with  the   degeneracy  between  the
measurement of line-of-sight velocity dispersion and $h_4$.

\end{abstract}


%
%

\section{Introduction}
\label{sec:intro}

Roughly one-quarter of the visible light emitted by stars in the local
universe comes  out of the bulges of  disk galaxies \citep{per_sal_92,
fuk_etal_98}.   Understanding how  bulges form  is therefore  of great
importance to developing a  complete picture of galaxy formation.  The
processes by  which bulges form are  still debated.  On  the one hand,
the merger of dwarf-sized galactic  subunits has been suggested as the
main path for bulge  formation \citep{kau_etal_93}, which is supported
by the  relatively homogeneous bulge stellar populations  of the Milky
Way   and  M31  \citep{ferreras_etal_03,   zoc_etal_03,  ste_etal_03}.
Bulges  formed   in  such  mergers  are   termed  `classical'  bulges.
Alternatively, bulges  may form via internal  `secular' processes such
as   bar-driven  gas   inflows,  bending   instabilities,   and  clump
instabilities  \citep{com_san_81, pfenni_84,  com_etal_90, pfe_fri_91,
rah_etal_91,         noguch_99,         imm_etal_04,        athana_05,
deb_etal_06}. Moreover, the  growth of the bulge out  of disk material
may also  be externally triggered by satellite  accretion during minor
merging events  \citep{agu_etal_01, eli_etal_06} or  also via external
`secular'  processes.  Evidence for  secular bulge  formation includes
the   near-exponential   bulge   light   profiles   \citep{and_san_94,
and_etal_95,   cou_etal_96,   dejong_96,   car_etal_98,   pri_etal_01,
car_etal_01, caroll_99, mac_etal_03},  a correlation between bulge and
disk   scale  lengths   \citep{dejong_96,   mac_etal_03,  agu_etal_05,
men_etal_08},   the  similar   colors  of   bulges  and   inner  disks
\citep{ter_etal_94, pel_bal_96, cou_etal_96, car_etal_07}, substantial
bulge  rotation \citep{kormen_93,  kor_etal_02}, and  the  presence of
B/P-shaped   bulges  in   $\sim   45\%$  of   edge-on  disk   galaxies
\citep{lut_etal_00}.  A review of secular `pseudo-bulge' formation and
evidence  for it can  be found  in \citet{kor_ken_04}.   Standard cold
dark matter cosmology predicts  that galaxies without classical bulges
should be  rare \citep{don_bur_04}.   Not only  are they  not rare
\citep[e.g.,][]{lau_etal_07},  they are  more  prevalent among  normal
galaxies than classical bulges, exacerbating the disagreement between
theory   and  observations   \citep{deb_etal_06}.   It   is  therefore
important to determine which bulges are of the classical versus pseudo
variety, and which are a mix of both.

$N$-body  simulations show  that barred  galaxies have  a  tendency to
develop   B/P  bulges   \citep{com_san_81,   com_etal_90,  pfe_fri_91,
rah_etal_91, deb_etal_06}.  The fraction  of edge-on galaxies with B/P
bulges is comparable to the  fraction of disk galaxies containing bars
\citep[$\sim  60\%$][]{esk_etal_00, mar_jog_07}  hinting that  the two
are related.  The presence of bars in edge-on galaxies with B/P bulges
has   been  established   by   the  kinematics   of   gas  and   stars
\citep{kui_mer_95,  mer_kui_99,  bur_fre_99,  bur_ath_99,  bur_ath_05,
veg_etal_97,   ath_bur_99,  chu_bur_04}.    However,   the  degeneracy
inherent in deprojecting edge-on  galaxies makes it difficult to study
other properties  of the host  galaxy.  For example,  simulations show
that a bar  can produce a B/P shape even if  a massive classical bulge
formed before the disk \citep{ath_mis_02, deb_etal_05}.  Understanding
the  relative importance  of classical  and pseudo-bulges  requires an
attempt at a cleaner separation  of bulges, bars and peanuts, which is
easiest to accomplish in less inclined systems. At least in two cases,
NGC~4442  \citep{bet_gal_94} and  NGC~7582 \citep{qui_etal_97},  a B/P
bulge can be recognized at inclination $i < 75\degrees$, which permits
also a direct identification of the bar.  \citet{deb_etal_05} proposed
a kinematic diagnostic of B/P bulges which properly works up to a disk
inclination of about $30^\circ$ namely  a double minimum in the fourth
moment,  $h_4$, of  the line-of-sight  velocity  distribution (LOSVD),
along the  major-axis of the bar.   These minima occur  because at the
location of the B/P shape,  the vertical density distribution of stars
becomes  broader,  which leads  to  a  double  minimum in  $z_4$,  the
fourth-order   Gauss-Hermite   moment    of   the   vertical   density
distribution.   The kinematic  moment $h_4$  is  then found  to be  an
excellent proxy for the unobservable $z_4$.  In contrast, the increase
in the  vertical scale-height does not produce  any distinct signature
of a  B/P bulge and  the vertical velocity dispersion,  $\sigma_z$, is
too strongly  dependent on the  radial density variation to  provide a
useful  B/P  bulge diagnostic.   \citet{deb_etal_06}  showed that  the
diagnostic continues to hold even when gas is present since this sinks
to a radius smaller than that of the B/P bulge.

In this Letter we present the first high resolution stellar kinematics
of face-on galaxies  with the goal of searching  for B/P-shaped bulges
and testing the diagnostic proposed by \citet{deb_etal_05}. In \S
\ref{sec:obsred} we  describe the observations of the  sample and data
reduction,    while    our     results    are    presented    in    \S
\ref{sec:res}.   Finally,   the    conclusions   are   given   in   \S
\ref{sec:conc}.

\section{Observations and data reduction}
\label{sec:obsred}

\subsection{Sample selection}

The barred  galaxies NGC~98  and  NGC~600 were  selected in  the
NASA/IPAC  Extragalactic Database  (NED) as  bright ($B_T  <  14$) and
undisturbed objects, with a disk inclination lower than $30^\circ$ and
a  bar length  larger than  $10''$ to  efficiently test  the kinematic
diagnostic, a disk diameter smaller than $4'$ to allow an accurate sky
subtraction, no strong evidence of dust and no bright foreground stars
in  the Digitized  Sky Survey  image. To  select the  sample galaxies,
their disk inclination, bar length,  and bar position angle were first
determined from  ellipse fits to the 2MASS  $J-$band archival images.
The  main  properties  of  the  sample galaxies  are  given  in  Table
\ref{tab:gals}.

\placetable{table1}

\subsection{Long-slit spectroscopy}  
\label{sec:spectroscopy}

The spectroscopic observations were carried out in service mode at the
Very Large Telescope (VLT)  at the European Southern Observatory (ESO)
on 2005 October 2-9 and November 2.
The Focal  Reducer Low Dispersion  Spectrograph 2 (FORS2)  mounted the
volume-phased  holographic  grism   GRIS\_1028z$+$29  with  1028  $\rm
grooves\;mm^{-1}$ and the $0\farcs7\,\times\,6\farcm8$ slit.
The detector was a mosaic of 2 MIT/LL CCDs, each with $2048\times2068$
pixels of $15\,\times\,15$ $\mu$m$^2$.  The wavelength range from 7681
to  9423 \AA\ was  covered in  the reduced  spectra with  a reciprocal
dispersion  of  0.858  \AA\   pixel$^{-1}$  and  a  spatial  scale  of
$0\farcs250$ pixel$^{-1}$ after a $2\times2$ pixel binning.
The  spectra were  taken  along the  bar  major axis  of NGC~98  ($\rm
P.A.=32\fdg2$) and NGC~600  ($\rm P.A. = 22\fdg1$) and  along the disk
major axis of NGC~1703 ($\rm P.A. = 148\fdg8$).  The total integration
time for  each galaxy  was 3  hours, in four  exposures of  45 minutes
each.
Comparison lamp  exposures obtained  for each observing  night ensured
accurate  wavelength calibrations.  Spectra  of G  and  K giant  stars
served as kinematic templates.  The average seeing FWHM was $1\farcs3$
for  NGC~98, $0\farcs9$ for  NGC~600, and  $0\farcs9$ for  NGC~1703 as
measured from the ESO Differential Imaging Meteo Monitor.

Using  standard IRAF\footnote{IRAF  is distributed  by NOAO,  which is
operated  by  AURA Inc.,  under  contract  with  the National  Science
Foundation.}   routines,  all  the   spectra  were   bias  subtracted,
flat-field  corrected,  cleaned  of  cosmic rays,  corrected  for  bad
pixels, and wavelength calibrated as in Debattista et al. (2002).  The
accuracy of the wavelength  rebinning ($\approx1 \kms$) was checked by
measuring wavelengths  of the brightest night-sky  emission lines. The
instrumental resolution was $1.84\pm0.01$ \AA\ (FWHM) corresponding to
$\sigma_{\rm inst} = 27$ $\kms$ at 8552 \AA .
The  spectra obtained  for the  same  galaxy were  co-added using  the
center of the stellar continuum as reference. In the resulting spectra
the  sky  contribution  was  determined  by  interpolating  along  the
outermost $\approx30''$ at the edges of the slit and then subtracted.

\subsection{Photometry}

We analyzed the uncalibrated acquisition images from the VLT to derive
the photometric properties of the sample galaxies.
Isophote-fitting with ellipses, after masking foreground stars and bad
pixels, was carried out using the IRAF task ELLIPSE.
Under  the  assumption  that  the  outer  disks  are  circular,  their
inclination  was determined by  averaging the  outer isophotes.  All 3
galaxies have $i < 30\degrees$ (Table~\ref{tab:gals}).
The semi-major axis length, $\len$,  of the bars of NGC~98 and NGC~600
(Table~\ref{tab:gals}) was measured from a Fourier decomposition as in
\citet{agu_etal_00}.   It was  calculated using  the bar/interbar
intensity ratio  $I_{\rm b}/I_{\rm ib}$. The bar  intensity is defined
as the sum of  the even Fourier components, $I_{0}+I_{2}+I_{4}+I_{6}$,
while the  inter-bar intensity is  given by $I_{0}-I_{2}+I_{4}-I_{6}$.
The  bar length is  the FWHM  of radial  profile of  $I_{\rm b}/I_{\rm
ib}$. This method was applied by \citet{ath_mis_02} to analytic models
demonstrating its accuracy in measuring $\len$.

\placefigure{f1}

\subsection{Kinematics}

The stellar  kinematics of the  three galaxies were measured  from the
galaxy absorption features present in the wavelength range centered on
the Ca~{\small  II} triplet ($\lambda\lambda\,8498,\,8542,\,8662\,$\AA
)      using     the      Penalized      Pixel     Fitting      method
\citep[pPXF,][]{cap_ems_04}.  The  spectra  were  rebinned  along  the
dispersion  direction to a  logarithmic scale,  and along  the spatial
direction   to   obtain  a   signal-to-noise   ratio   $S/N  \ga   40$
\AA$^{-1}$.  It  decreases to  $S/N=20$  \AA$^{-1}$  at the  outermost
radii.

A linear  combination of the template stellar  spectra, convolved with
the  line-of-sight  velocity  distribution  (LOSVD) described  by  the
Gauss-Hermite  expansion  by  \citet{vdm_fra_93}  was fitted  to  each
galaxy spectrum by $\chi^2$  minimization in pixel space. This allowed
us to derive  profiles of the mean velocity  ($v_{\rm los}$), velocity
dispersion  ($\sigma_{\rm  los}$),  third  ($h_3$),  and  fourth-order
($h_4$)  Gauss-Hermite  moments. The  uncertainties  on the  kinematic
parameters were estimated by Monte Carlo simulations with photon, read
out and sky noise.
Extensive testing on simulated galaxy spectra was performed to provide
an  estimate  of  the biases  of  the  pPXF  method with  the  adopted
instrumental setup  and spectral sampling. The  simulated spectra were
obtained by convolving the  template spectra with a LOSVD parametrized
as a Gauss-Hermite  series and measured as if they  were real. No bias
was  found  in  the  ranges  of $S/N$  and  $\sigma_{\rm  los}$  which
characterize the spectra  of the sample galaxies. The  values of $h_3$
and $h_4$ measured for the simulated spectra differ from the intrinsic
ones only within the measured errors \citep[see also][]{ems_etal_04}.

\section{Results}
\label{sec:res}

\begin{figure*}[ht!]
\begin{center}
\includegraphics[width=2.2in, bb=110 350 460 1100]{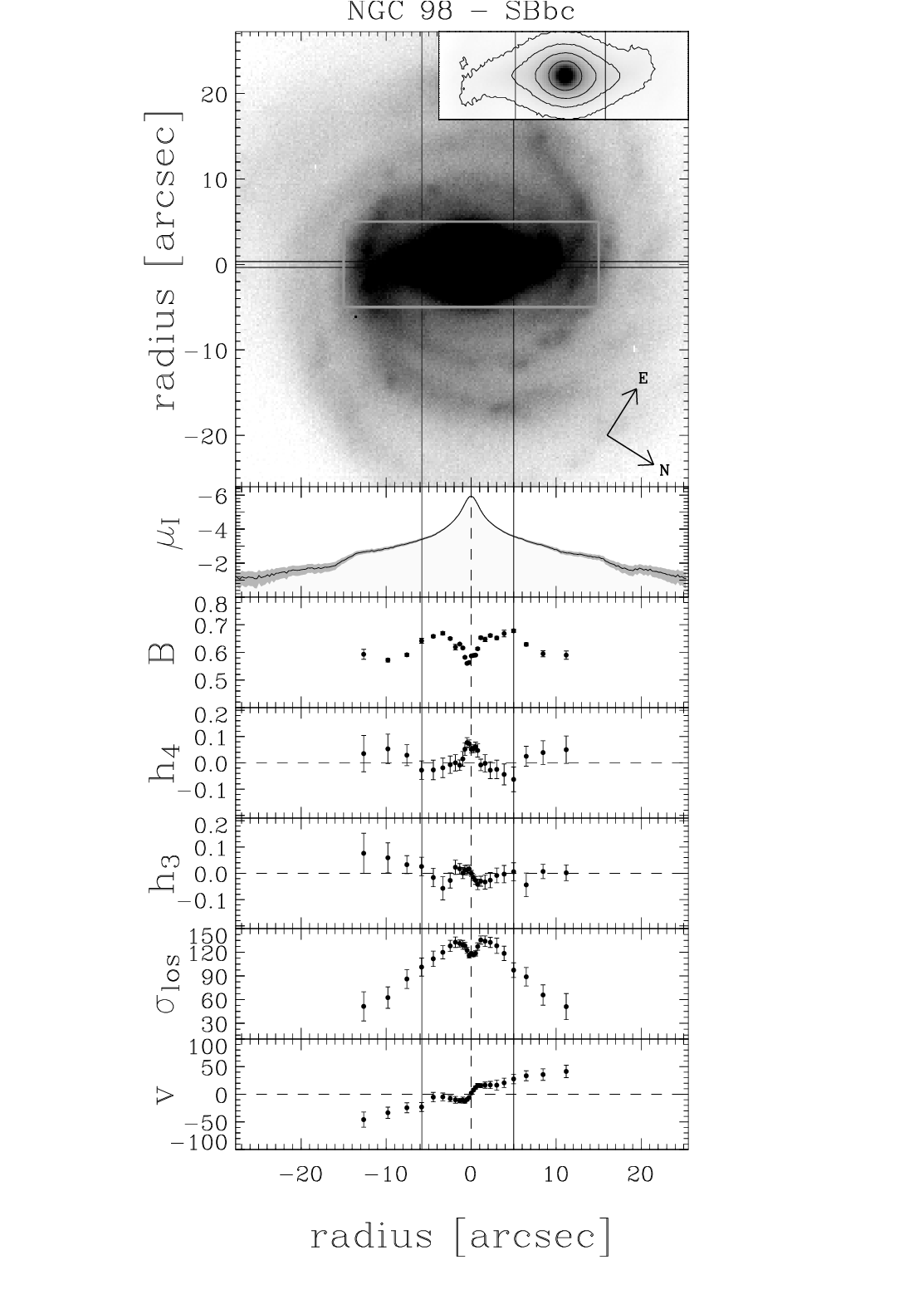}
\includegraphics[width=2.2in, bb=110 350 460 1100]{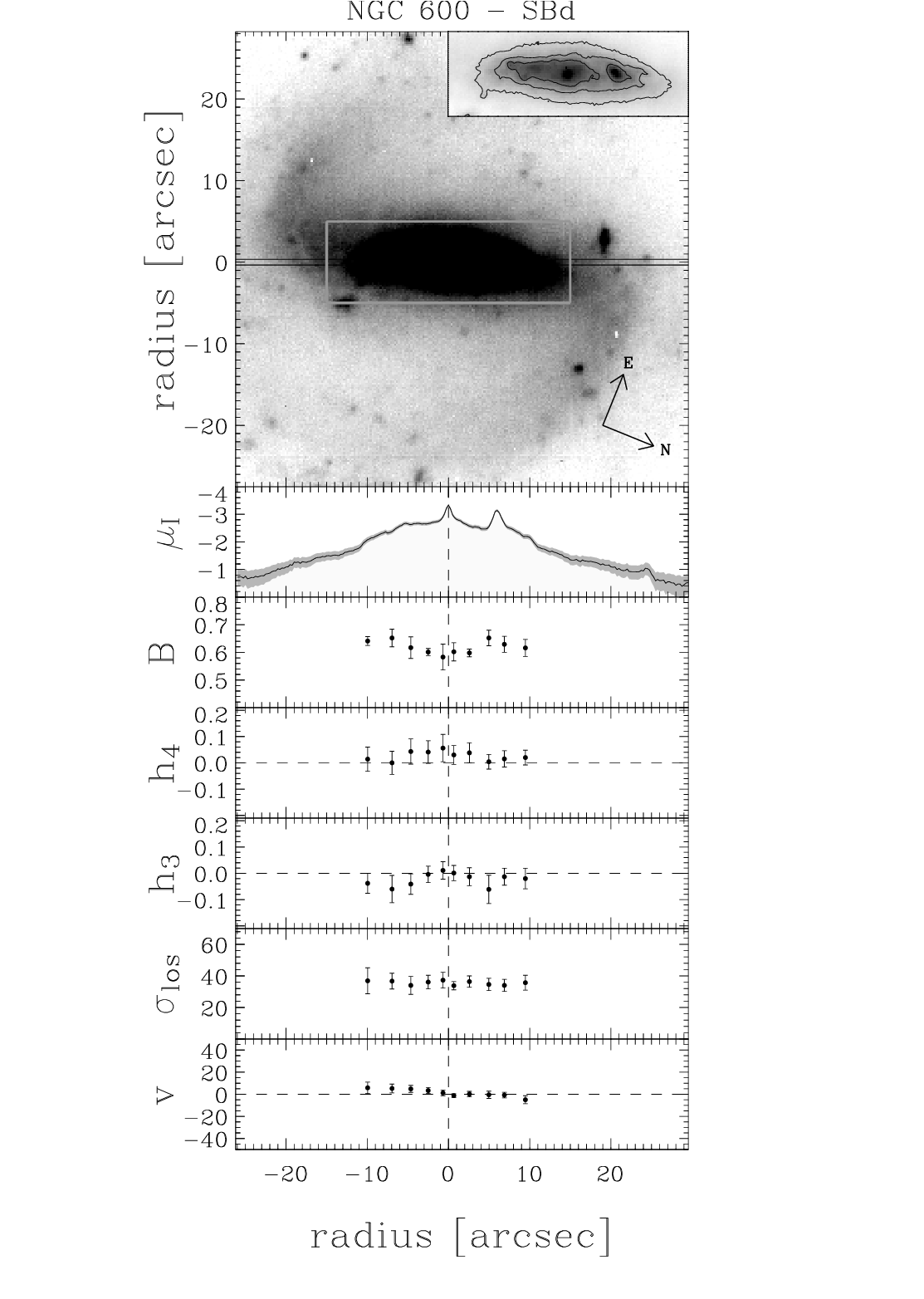}
\includegraphics[width=2.2in, bb=110 350 460 1100]{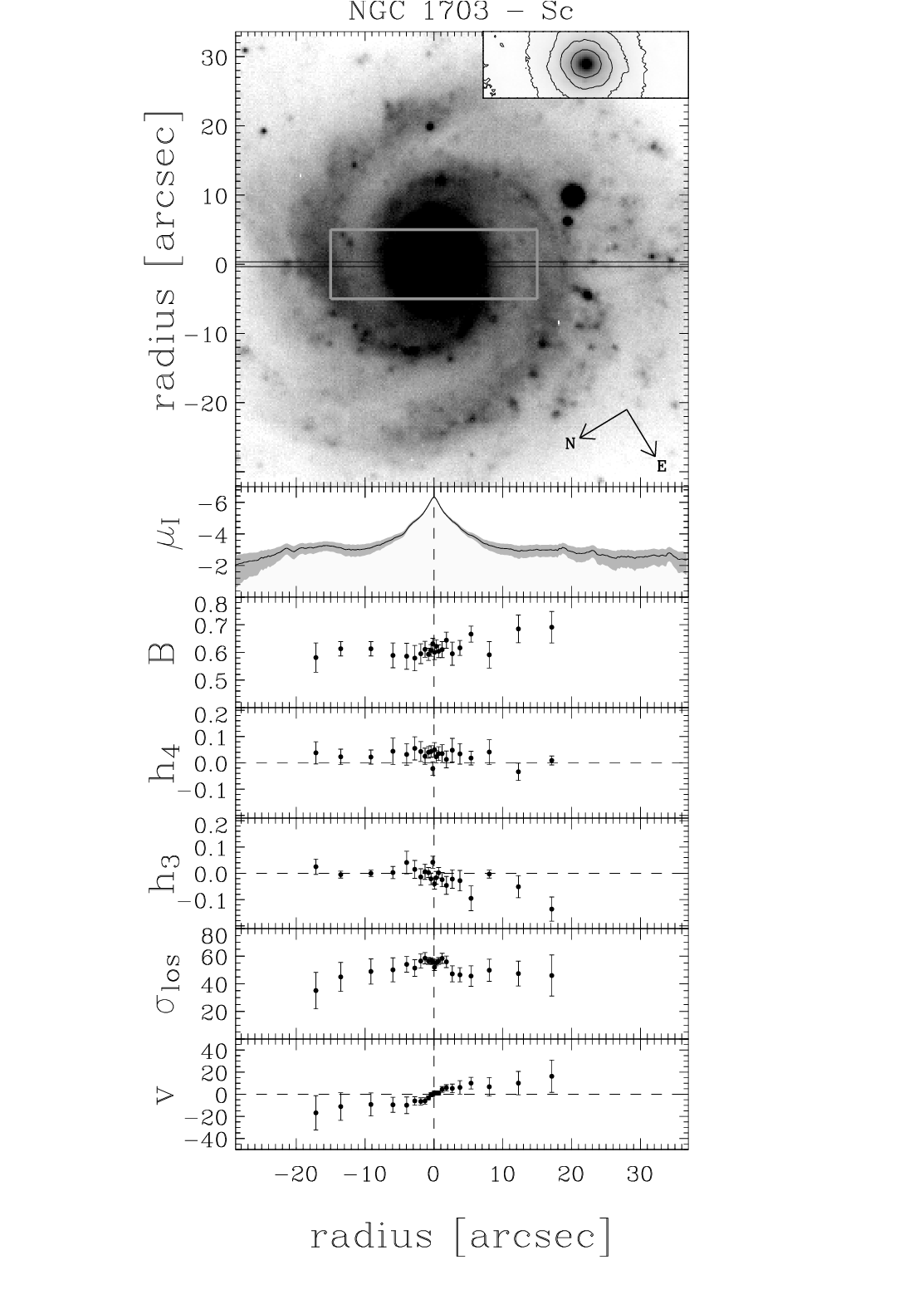}
\end{center}
\caption{Morphology and stellar kinematics of NGC~98 (left panels),
 NGC~600 (central panels), and NGC~1703 (right panels).  
 For each galaxy the top panel shows the VLT/FORS2 $I-$band image.
 The slit position and image orientation are indicated. The inset
 shows the portion of the galaxy image marked with a white box. The
 gray scale and isophotes were chosen to enhance the features observed
 in the central regions.
 The remaining panels show from top to bottom the radial profiles of
 surface brightness (extracted along the slit with an arbitrary zero
 point), ${\cal B}(0.7,0.9)$, $h_4$, $h_3$, $\sigma_{\rm los}$ (in
 $\kms$ ), and velocity $v$ (in $\kms$ , obtained by subtracting the
 systemic velocity from $v_{\rm los}$). The two vertical lines
 indicate the location of the $h_4$ minima in NGC~98.}
\label{fig:kine}
\end{figure*}

The photometric and  kinematic profiles of the sample  of galaxies are
shown in Figure~\ref{fig:kine}. We focus here on the $h_4$ profiles; a
full  analysis of this  data set  will be  presented in  a forthcoming
paper.  For the unbarred galaxy,  NGC~1703, the $h_4$ profile shows no
sign of a minimum.  The same  is true for the almost bulgeless, barred
NGC~600.  In both these galaxies  $\sigma > 30 \kms$, including within
the bar in NGC~600. Thus, our  failure to find an $h_4$ minimum cannot
be ascribed to low spectral  resolution. In NGC~98, instead, we find a
double minimum in the $h_4$ profile. The two minima are symmetric with
respect to the galaxy center ($|r|\approx5''$).  In the simulations of
\citet{deb_etal_05} the  minima in $h_4$  are $\sim 0.05 -  0.1$ deep,
comparable to the minimum in the profile of NGC~98.

We  confirmed that  our kinematics,  and minima  in $h_4$,  are  not a
result of details  of the pPXF method by  repeating the analysis using
the  Fourier Correlation  Quotient (FCQ)  method  \citep{bender_90} as
done in \citet{piz_etal_08}.   Nonetheless expansions in Gauss-Hermite
moments  are degenerate \citep{gerhar_93,  vdm_fra_93} since,  for the
even moments, it is possible to obtain comparably good fits by trading
$\sigma_{\rm los}$ for $h_4$.  In  order to confirm that the signature
of a B/P bulge  obtained in NGC~98 is not due to  a conspiracy of such
degeneracies  we introduced  an LOSVD  broadening measure  defined for
each radius as
\begin{equation} 
{\cal B}(x_1,x_2) \equiv W(x_2)/W(x_1),
\end{equation}
where 
\begin{equation}
W(x) = \int_{v^{\rm min}_{\rm los}}^{v^{\rm max}_{\rm los}} F(v)dv,  
\end{equation}
with $F(v)$  the cumulative distribution  of the LOSVD. The  values of
$v^{\rm  min}_{\rm los}$  and  $v^{\rm max}_{\rm  los}$ correspond  to
$F(v^{\rm min}_{\rm los}) =  0.5\,(1-x)$ and $F(v^{\rm max}_{\rm los})
= 0.5\,(1+x)$, respectively.
${\cal B}(x_1,x_2)$  has the advantage  of not being sensitive  to the
wings of the distribution, unlike the kurtosis and is a non-parametric
measure of  the LOSVD  shape.  Experiments with  $N$-body simulations,
including those  presented in \citet{deb_etal_05},  showed that ${\cal
B}(0.7,0.9)$  is a  useful peanut  diagnostic  in the  same spirit  as
$h_4$.  Figure \ref{fig:kine} plots ${\cal B}(0.7,0.9)$.  A clear peak
in the broadening is apparent  at $|r| \simeq 5\arcsec$ in NGC~98.  We
conclude that the evidence for a B/P-shaped bulge in NGC~98 is robust.

\section{Conclusions}
\label{sec:conc}

We have identified  a B/P-shaped bulge in one galaxy  from a sample of
three face-on  galaxies.  In the  unbarred control galaxy  NGC~1703 we
had  not  expected  to find  a  B/P  bulge.   The  failure to  find  a
double-minimum in  $h_4$ is  therefore fully consistent  with previous
results \citep{chu_bur_04}.   Of the  two barred galaxies,  NGC~98 has
clear evidence of a B/P bulge  while NGC~600 does not.  The absence of
a B/P shape in the latter galaxy is not surprising since it appears to
not have a bulge and its bar has a lumpy structure.

If we identify the radius of  the B/P bulge, $\rbp$, with the location
of the minimum in  $h_4$, as found in simulations \citep{deb_etal_05},
then  we  find  $\rbp \simeq  0.35  \len$,  where  $\len$ is  the  bar
semi-major axis.  Similarly, \citet{kor_ken_04} noted that the maximum
radius  of  the boxy  bulge  is about  one-third  of  the bar  radius.
Simulations also  produce B/P-bulges which are  generally smaller than
the bar \citep{she_sel_04, deb_etal_05}.

While  observations  have established  the  presence  of  bars in  the
majority  of  systems  with  B/P-bulges,  the converse  has  not  been
established.   \citet{deb_etal_05} demonstrate how  the presence  of a
classical  bulge can in  some cases  mask the  presence of  B/P bulge.
Moreover, the ratio of radii of bulges and B/P bulges is determined by
resonances.  For these various reasons it would be very instructive to
repeat measurements such as here for a sample of barred galaxies.


%
\acknowledgements We thank Eva Grebel for her support and Lorenzo Morelli for
fruitful discussion.
E.M.C. and V.P.D. thank the Instituto de Astrof\'\i sica de Canarias for
hospitality during part of this project.
J.M.A. acknowledges support from the Istituto nazionale di Astrofisica (INAF).
E.M.C. and A.P. receive support from the grant CPDA068415/06 by Padua University.
S.D.R. acknowledges Postdoctoral Fellowship support from the Fund for 
Scientific Research--Flanders, Belgium (FWO).
J.A.L.A. is funded by the Spanish DGES, grant AYA-2007-67965-C03-01.
This research has made use of the NASA/IPAC Extragalactic Database,
Digitized Sky Survey, and 2 Micron All Sky Survey.


\bibliographystyle{aj.bst}
\bibliography{allrefs}

\clearpage

\begin{deluxetable}{llcccrc}
\tablecolumns{7}
\tablewidth{0pc}
\tablecaption{Parameters of the sample galaxies.}
\tablehead{
\multicolumn{1}{c}{Object} &
\multicolumn{1}{c}{Type} &
\multicolumn{1}{c}{$B_T$} &
\multicolumn{1}{c}{$D_{25}$} &
\multicolumn{1}{c}{$M^0_{B_{T}}$} &
\multicolumn{1}{c}{$i$} &
\multicolumn{1}{c}{$a_B$}}
\startdata
NGC~98   & SBbc & 13.4 & $1\farcm7$ & $-22.4$ & $26^\circ$ & $14\arcsec \pm 1\arcsec$ \\
NGC~600  & SBd  & 12.9 & $3\farcm3$ & $-20.0$ & $21^\circ$ & $12\arcsec \pm 1\arcsec$ \\
NGC~1703 & Sc   & 11.9 & $3\farcm0$ & $-20.6$ & $15^\circ$ & \nodata 
\enddata
\tablecomments{Hubble types, apparent magnitudes, and diameters 
are from the NASA/IPAC Extragalactic Database (NED). Absolute
magnitudes are calculated from $B^0_T$ (NED) with distance from
$V_{\rm CMB}$ (NED) and $H_0=75$ $\kmskpc$. Disk inclinations and bar
semi-major axis lengths are from this paper.}
\label{tab:gals}
\end{deluxetable}

\clearpage

\end{document}